\documentclass[oneside,12pt]{article}
\usepackage{setspace}
\usepackage[utf8]{inputenc}
\usepackage{authblk}
\usepackage{graphicx,bm,color}
\usepackage[dvipsnames]{xcolor}
\usepackage[margin=1in]{geometry}
\usepackage{amsthm,amsmath,amssymb,mathtools}
\usepackage{multirow}
\usepackage{todonotes}
\usepackage{tikz-cd} % for commutative diagrams
\usepackage{authblk}

\definecolor{britishracinggreen}{rgb}{0.0, 0.5, 0.15}
\definecolor{burgundy}{rgb}{0.5, 0.0, 0.13}
\definecolor{egyptianblue}{rgb}{0.06, 0.2, 0.65}
\usepackage[colorlinks=true, citecolor=burgundy, linkcolor=britishracinggreen, urlcolor=egyptianblue]{hyperref}

\definecolor{britishracinggreen}{rgb}{0.0, 0.5, 0.15}
\definecolor{burgundy}{rgb}{0.5, 0.0, 0.13}
\definecolor{egyptianblue}{rgb}{0.06, 0.2, 0.65}
\usepackage[colorlinks=true, citecolor=burgundy, linkcolor=britishracinggreen, urlcolor=egyptianblue]{hyperref}

\usepackage{braket}

\newcommand{\ketbra}[2]{\mbox{$|#1\rangle\langle #2|$}}

\def\eye{\mathbb{1}}
\def\herm{\textrm{herm}}
\def\spec{\textrm{spec}}
\def\Mat{\textrm{Mat}}

\def\ketbra#1#2{{\vert#1\rangle\!\langle#2\vert}}
\def\braket#1#2{{\langle#1\vert#2\rangle}}
\usepackage{slashed}
\def\eye{\mathbb{I}} %added a macro here will organize later 

\DeclareMathOperator{\Span}{span}

\newtheorem{theorem}{Theorem}

\newtheorem{corollary}{Corollary}
\newtheorem{example}{Example}
\newtheorem{remark}[theorem]{Remark}
\newtheorem{definition}[theorem]{Definition}

\graphicspath{{./pics/}}

\begin{document}

\author[1, 2]{Jacob Biamonte\footnote{Parts of this work were completed at the Simons Institute for the Theory of Computing, at the University of California, Berkeley.}}

\affil[1]{\footnotesize NASA Quantum Artificial Intelligence Laboratory, NASA Ames Research Center, Moffett Field, CA 94035, USA}
\affil[2]{\footnotesize  Research Institute for Advanced Computer Science, University Space Research Association, NASA Ames Research Center, Moffett Field, CA 94035, USA}

\title{On Commutative Penalty Functions in Parent-Hamiltonian Constructions}

\date{\today}

\maketitle
\begin{abstract}
There are several known techniques to construct a Hamiltonian with an expected value that is minimized uniquely by a given quantum state. Common approaches include the parent Hamiltonian construction from matrix product states, building approximate ground state projectors, and, in a common case, developing penalty functions from the generalized Ising model. Here we consider the framework that enables one to engineer exact parent Hamiltonians from commuting polynomials. We derive elementary classification results of quadratic Ising parent Hamiltonians and to generally derive a non-injective parent Hamiltonian construction. We also consider that any $n$-qubit stabilizer state has a commutative parent Hamiltonian with $n+1$ terms and we develop an approach that allows the derivation of parent Hamiltonians by composition of network elements that embed the truth tables of discrete functions into a kernel space. This work presents a unifying framework that captures components of what is known about exact parent Hamiltonians and bridges a few techniques across the domains that are concerned with such constructions.
\end{abstract}

% \tableofcontents 

% \listoftodos 

\newpage 
\section{Introduction}

We are concerned with engineering the ground space of Hamiltonian operators using systematic mathematical methods.  This has several motivations.  

In models of Hamiltonian quantum computation, a Hamiltonian is programmed so that its lowest energy state stores the solution to a given problem instance \cite{Kempe2006ComplexityLocalHamiltonian, Lucas2014Ising, BL08, WFB12}.  One then has to minimize the Hamiltonian to solve the problem instance.  Ground state models of quantum computation turn out to relate to several naturally occurring physical processes that can be bootstrapped to form a computational process which naturally minimizes a system's Hamiltonian. 

This same question arises in the area of quantum spin-chain physics.  Given a known physical quantum state, what Hamiltonian governs its dynamics?  The AKLT model \cite{Affleck1987ValenceBond}, named after physicists Affleck, Kennedy, Lieb, and Tasaki, is a notable example in quantum physics that illustrates the relationship between a specific quantum state and its corresponding parent Hamiltonian \cite{Smith2023Deterministic}. Generally, the field of matrix product states have developed a theory of parent Hamiltonains which encompass 1D spin physics \cite{Per+06, Fern_ndez_Gonz_lez_2014}.  The theory of invective projected entangled pair states (PEPs) provides the correspondence between parent Hamiltonains and 2D tensor networks \cite{PerezGarcia2007PEPS}.    

In the area of adiabatic quantum computation, one considers diagonal operators arising from the Ising model as well as universal non-commutative models \cite{BL08}.  Various methods have been developed to control penalty functions.  This includes various methods that embed classical logic \cite{WFB12}, and also graph problems as well as the 21 NP-complete problems from Karp \cite{Lucas2014Ising}.  

In variational quantum computation, the variational quantum eigensolver is a general way to minimize quantum Hamiltonians.  In this setting, several interesting problem Hamiltonian constructions exist, such as those based on the Feynman-Kiteav construction, but other constructions also exist \cite{Bia21}.  

Generally speaking, the parent Hamiltonian problem is also related to the area of downfolding or effective Hamiltonian theory.  Such methods have arisen in quantum information science in the area of perturbative gadgets \cite{Kempe2006ComplexityLocalHamiltonian, Cao2015Hamiltonian}.  In these methods, one relies on a 2-body operator to approximate a higher order operator by introducing slack bits which decouple in the engineered coupling regime \cite{Kempe2006ComplexityLocalHamiltonian, Cao2015Hamiltonian, BL08}.

We are concerned with the special class of commutative parent Hamiltonian and we consider a merger of techniques \cite{WFB12, Biamonte2008, Bia21, Biamonte2013TensorNetwork, Bergholm2011Categorical, Sengupta2022TensorNetworks}  and approaches to develop a framework which we are applying to better understand non-perturbative and algebraic parent Hamiltonian constructions more generally.  

\newpage 

Given the description of a quantum state $\ket{\psi}\in [\mathbb{C}_d]^{\otimes n}$ of $n$-interacting, $d$-level systems, a parent Hamiltonian for $\ket{\psi}$ is a Hermitian operator on $[\mathbb{C}_d]^{\otimes n}$ that has a non-negative spectrum and satisfies: 
\begin{equation}
    H\ket{\psi} = 0.  
\end{equation}
Formally: 
\begin{definition}[Parent Hamiltonian]
    Consider a unit vector $\ket{\psi} \in [\mathbb{C}_d]^{\otimes n}$ and a non-negative Hamiltonian $H \in \herm_\mathbb{C}(d)$.   When $H$ is chosen so that $H \ket{\psi} = 0$, $H$ is the parent Hamiltonian of $\ket{\psi}$.
\end{definition}
The operator $H$ is constrained to have desirable properties such as: 
\begin{enumerate}
    \item (Algebraic locality) To limit the degree of the Hamiltonian, e.g.~Hamiltonians with only 1- and 2-body interaction terms \cite{Kempe2006ComplexityLocalHamiltonian, Lucas2014Ising, BL08, WFB12}. 
    \item (Spatial locality) To limit (also) the distance between Hamiltonian terms which connect interacting particles.  For example, limiting to nearest and next nearest neighbor interactions \cite{Per+06, Fern_ndez_Gonz_lez_2014}.  
    \item (Operator cardinality) To limit the number of non-trivial terms an operator has expressed in a local (Pauli) basis \cite{Kempe2006ComplexityLocalHamiltonian, Bia21}. 
    \item (Uniqueness) To limit the dimension of the kernel of $H$ so that only vectors in $\Span{\ket{\psi}}$ minimize $H$ \cite{Per+06, Fern_ndez_Gonz_lez_2014, PerezGarcia2007PEPS, BL08, WFB12}.  
\end{enumerate}
Further restrictions are also possible.  

Several parent Hamiltonian constructions have been discovered.  There is the well known construction appearing in injective matrix product states (MPS) and similarly injective projectile entangled pair states (PEPS) \cite{Per+06}. Perhaps less known, the so-called `inverse Ising problem' seeks to determine a quadratic Ising model that is minimized on elements of a data set \cite{NZB17}.  The case of Ising model reductions have also become common place in quantum information processing \cite{WFB12, Biamonte2008, H21} and rely heavily on various aspects of Boolean algebra \cite{Odn14}.  In the setting of quantum computing, we are also familiar with the Feynman-Kiteav construction: here a given quantum circuit is mapped to a Hamiltonian that has limited algebraic (though not spatial) locality.  This construction has been used in the universality proofs of the adiabatic \cite{Aha+08, BL08} and variational \cite{Bia21} models of quantum computation.  

The present work is part of the ongoing effort to understand how to engineer the ground states of Hamiltonians.  After recalling the preliminary definitions next in \S~\ref{sec:framework}, we continue on to consider the basic properties of commutative parent Hamiltonians in \S~\ref{sec:parent}.  We apply a fairly standard construction to the create a non-degenerate but also $n$-parent Hamiltonian of the $\ket{\text{GHZ}}$ state.  Similarly, the parent Hamiltonian of any stabilizer state can be constructed with has not more than $n+1$ terms.  We carry on to define some of the basic elements of this theory: the commutative cone, as well as its variant with tensor structure and its restriction to diagonal wedges.  We also define the kernel algebra for a diagonal $\otimes$-cone in Definitions \ref{def:kernelalgebra1} and \ref{def:kernelalgebra2}.  We then change focus towards the generalized Ising model and pinpoint its relationship with operator and state embeddings in \S~\ref{sec:pseduo}.  This includes the classification of elementary parent Hamiltonians in \S~\ref{sec:classification}.  In particular we consider the analysis of one-body forms and use this analysis to construct quadratic forms.  In this setting, Equation \eqref{eq:ghzisingparent} provides an alternative GHZ parent Hamiltonian construction that is degenerate and also uses an all-to-all two-body interaction.  From this we turn to the root system of symmetric polynomials and develop in some detail a few examples.  Using the developed framework we finally turn to embedding and composing discrete functions in the kernels of diagonal $\otimes$-cones.  We consider an elementary non-injective Parent Hamiltonian construction and also how to construct the parent Hamiltonian of any given set of bit strings, provided we have access to some number of slack bits that grows directly with the number of operations in its network form.  

\section{Framework}\label{sec:framework}

We will describe the basic framework used in this study.  In quantum mechanics, a Hamiltonian describes the evolution of a quantum system over time. In our setting, we fix a basis and express the Hamiltonian as a matrix of finite dimension. Formally: 

\begin{definition}[Hamiltonian]
A Hamiltonian $H$ is represented as a Hermitian operator ($H=H^\dagger$) on a complex vector space, viz. 
\begin{equation}
    H\in \herm_{\mathbb{C}}(d)\equiv \{ M\in \Mat_{\mathbb{C}}(d)|~ M=M^\dagger \}
\end{equation}
$\forall l$, $\lambda_l \in \spec(H), ~\lambda_l \in \mathbb{R}$. We call $\lambda$ an energy, where the ground state energy is,
\begin{equation}
    \lambda_0(H)= \min_{\ket{\psi}\in\mathbb{C}^d} \frac{\langle \psi |H| \psi \rangle}{\braket{\psi}{\psi}}. 
\end{equation}
\end{definition}
\begin{remark} [Notation]
    We use $\lambda_l(H)$ to denote the $l$th eigenvalue of $H$.     We write $H\geq 0$ to mean  $\forall l, \lambda_l(H)\geq 0 \iff \forall \ket{\psi}, \bra{\psi}H\ket{\psi}\geq 0$.  We call $H$ non-negative (positive semidefinite is common in quantum physics literature). 
\end{remark}

\begin{definition}[Pauli matrices]
    We consider the standard Pauli matrices, \(X\), \(Y\), \(Z\), which satisfy \(XYZ = \imath\eye\), where each matrix squares to the identity.
\end{definition}

\begin{definition}[Pauli cardinality]
    The Pauli cardinality is the number of terms in the Pauli basis expansion of qubit Hamiltonian $H$, written $|H|_c$. 
\end{definition}

\begin{definition}[Albegraic \( k \)-locality]
Let \( \mathcal{L}(\mathbb{C}_d) \) be the space of linear operators on \( \mathbb{C}_d \), with a Hermitian basis given by \( \Lambda_l \) for \( l \in \{ 1,\ldots,d^2 \} \). Consider a Hamiltonian on \( n \) \( d \)-level systems expanded with real numbers $c_s$ over this basis. We represent the Hamiltonian as:
\begin{equation}
H = \sum_{s\in\{1:d^2\}^{\times n}} c_s \bigotimes_{j=1}^n \Lambda_{s_j}
\end{equation}
where \( s_j \) is the \( j^{th} \) element of the tuple \( s \). The Hamiltonian \( H \) is said to be \( k \)-local if \( k \) is the maximum number of non-identity terms in any tensor product term with a non-zero coefficient \( c_s \).
\end{definition}

\begin{example}[Algebraic locality]
The operator \( \eye^{\otimes(n-2)} \otimes X \otimes X \) is \( 2 \)-local since it acts non-trivially on two \( d \)-level systems.
\end{example}

\begin{example}[Spatial locality]
The \( 2 \)-local operator \( X \otimes \eye^{\otimes(n-2)} \otimes X \) acting on a line of $n$ qubits has spatial locality $n-2$.  
\end{example}

\begin{remark}[Notation]
    $|\cdot |$ denotes the cardinality of a set.
\end{remark}

\begin{definition}[Gapped Hamiltonian]
If $|\arg\min_{\ket{\psi}\in\mathbb{C}^d}\langle \psi |H| \psi \rangle|=1$, then the following are equivalent:
\begin{enumerate}
\item $H$ is gapped.
\item $H$ has a non-degenerate ground space.
\item $\arg\min_{\ket{\psi}\in\mathbb{C}^d}\langle \psi |H| \psi \rangle= \Span_{\mathbb{C}} \{ \ket{\lambda_0} \}\cong \mathbb{C}$.
\end{enumerate}
\end{definition}

\begin{remark}[Notation]
Let $\{\ket{b_l}\}$ be a basis of $\mathbb{C}_s$. Hence, $1 \leq l \leq s$ for $s\in \mathbb{N}$, then $\ket{b_1 \cdots b_d}$ forms a basis for $[\mathbb{C}_s]^{\otimes d}$. We write $\ket{b_1 \cdots b_d}$ for $b_1 \cdots b_d := \bf{b}$. 
\end{remark}

\begin{remark}[Notation]
    We use $\langle \{S\} \rangle = \Span \{S\}$ to denote the $\mathbb{C}$-linear extension of the set S. 
\end{remark}

\subsection{Parent Hamiltonians}\label{sec:parent} 

The following example establishes that a parent Hamiltonian always exists.  This construction generally requires an exponential number of local terms when expressed in a tensor product basis, such as the Pauli basis.  

\begin{example}[Existence by the trivial construction]\label{ex:trivial}
Let $\Gamma = \eye - \ketbra{\psi}{\psi}$ then $\Gamma$ is Hermitian and $\Gamma^2=\Gamma$.  We have that $\Gamma \ket{\psi} = \lambda_\psi \ket{\psi} = \lambda_\psi^2 \ket{\psi}$ 
which has solutions only over $\mathbb{B}$ which implies that $\Gamma \geq 0$.
\end{example}

\begin{example}[Exact ground state projectors \cite{Aha+11, Ara+12, Ara+13}]\label{ref:agsp}
Given $\ket{\psi} \in \mathbb{C}_d$ such that 
\begin{equation}
    \ket{\psi} = U\ket{0^n}.  
\end{equation}
Then the sum of the individual projectors $\ket{1}\bra{1}^{(l)}$ acting on the $l$th spin is: 
\begin{equation}\label{eq:tele}
    H = \sum_{l=1}^n U\ket{1}\bra{1}^{(l)}U^\dagger.  
\end{equation}
The terms in the Hamiltonian commute and 
$\text{spec}(U\ket{1}\bra{1}^{(l)}U^\dagger)\in \{0, 1\}$. Equation \eqref{eq:tele} satisfies the eigenvalue equation $H U\ket{{\bf x}} = |{\bf x}|_1 U \ket{{\bf x}}$ where $|{\bf x}|_1$ is the one norm of bit string ${\bf x}$. 
\end{example}

The first construction (Example \ref{ex:trivial}) lacks practical utility due to the number of terms required to expand Equation \eqref{eq:tele} in the Pauli basis.  The second construction (Example \ref{ref:agsp}) will be considered in more detail.  Several works have relied partly on this construction to make approximate ground state projectors \cite{Aha+11, Ara+12, Ara+13}.  Likewise, the Pauli cardinality of Equation \eqref{eq:tele} is invariant whenever $U$ is a Clifford gate \cite{Bia21}. Let us construct the parent Hamiltonian of the $n$ qubit GHZ state by setting $U$ in Equation \eqref{eq:tele} to be a circuit preparing $\ket{\text{GHZ}_n} = \frac{1}{2^{n/2}}(\ket{0^n} +\ket{1^n})$.  The circuit preparing $\ket{\text{GHZ}_n}$ is well known:
\begin{equation}
    \label{eqn:GHZ_ckt}
    \ket{\text{GHZ}_n} = \prod_{j=1}^{n-1} (\text{CNOT})_{j, j+1} \text{H}_1 \ket{0^n}
\end{equation}
Here $(\text{CNOT})_{j,j+1}$ is a controlled not gate which acts on the $j+1$-th qubit with the control on the $j$-th qubit and $\text{H}_1$ is the Hadamard gate acting on the first qubit. One can now construct the parent Hamiltonian of $\ket{\text{GHZ}_n}$ for integer $n>1$ as:
\begin{equation}\label{eq:GHZ-parent}
    H_{\text{parent}} = \frac{1}{2} \left(n {\eye} - \sum_{j=1}^{n-1} Z_j Z_{j+1} - \bigotimes_{j=1}^n X_j\right). 
\end{equation}

$H_{\text{parent}}$ acts on bit string $\bf x$ as
\begin{equation}\label{eq:booleanaction}
    H_{\text{parent}}\ket{{\bf x}} = \left(\frac{n}{2} - \frac{1}{2} \left(\sum_{k=1}^{n-1} (-1)^{x_l + x_{l+1}}\right)\right) \ket{{\bf x}} - \frac{1}{2} \ket{\overline{{\bf x}}}
\end{equation}
where $\overline{{\bf x}}$ is the bit wise negation of $\bf x$. 
From Equation \eqref{eq:booleanaction} we arrive at the relation: 
\begin{equation}
    H_{\text{parent}}\ket{0^n} = - H_{\text{parent}}\ket{1^n}
\end{equation}
and so it immediately follows that the $n$-qubit GHZ state is in the kernel of the non-negative Hamiltonian $H_{\text{parent}}$. It can also be shown that $\text{ker}(H_{\text{parent}}) = \Span\{ \ket{\text{GHZ}_n}\}$. Breaking the sum in Equation \eqref{eq:booleanaction} into two non-negative parts, we let $|{\bf x}|_1$ be the Hamming weight of ${\bf x}$ and calculate: 
\begin{equation}
    \text{ker}\left(\frac{\eye - \bigotimes_{k=1}^n X_k}{2}\right) = \Span\{H^{\otimes n}\ket{{\bf x}}|~|{\bf x}|_1 \equiv 0 \pmod 2\}, 
\end{equation}
\begin{equation}
    \text{ker}\left( \frac{1}{2} \sum_{k=1}^{n-1} (\eye - Z_kZ_{k+1}) \right) = \Span\{\ket{0^n}, \ket{1^n}\}.  
\end{equation}
Both operators factor as $\ker \bigoplus \ker^\perp$ and so we are looking for normalized vectors that have non-trivial overlap with both kernels.  These are given by solutions to 
\begin{equation}
    \sum_{|{\bf x}|_1 ~\equiv ~0 \pmod 2}c_{\bf x} H^{\otimes n}\ket{{\bf x}} = a_0 \ket{1^n}+a_1\ket{0^n}.  
\end{equation}
From this we arrive at the only non-trivial solution ($a_0=a_1\neq 0$) which constrains the kernel as $\Span\{\ket{0^n}+\ket{1^n}\}$.  Given a corresponding Clifford circuit, a parent Hamiltonian with $n+1$ Pauli terms can be constructed for any stabilizer state. 

\subsection{Commutative $\otimes$-cone}
For any Hamiltonian $H$, there exists a constant $c$ such that the spectrum of $c\eye +H$ is non-negative.  Whenever we have $A, B \in \herm_{\mathbb{C}}(d)$, $A,B\geq 0$, then
\begin{equation}
     A+B\geq 0
\end{equation}
 and 
\begin{equation}
     c_1 A + c_2 B \geq 0
\end{equation}
$\forall c_1, c_2 \in \mathbb{R}_+$, where $\mathbb{R}_+$ is $[0, +\infty)$. This senario is called a cone or a tensor cone ($\otimes$-cone) if the underlying operators have tensor product structure. 

\begin{definition}[$\otimes$-cone] 
$\Span_{\mathbb{R}^+}\{ M\in \Mat_\mathbb{C}(d^n) |~ M\geq 0 \}$.  
\end{definition}

\begin{definition}[Commutative cone]
A cone $W$ where all elements $\Gamma_m$, $\Gamma_l$ in $W$ commute $\Gamma_l \Gamma_m = \Gamma_m \Gamma_l$ is a commutative cone. 
\end{definition}

\begin{definition}[Kernel]
Let $\Gamma$ be a square matrix.  The kernel or null space of $\Gamma$ is given as $\ker\{\Gamma\} = \Span_\mathbb{C}\{ \ket{\psi} |~  \Gamma \ket{\psi} = 0 \}$. 
\end{definition}

\begin{definition}[Union and intersection of kernels]
Let $A_k$ be a basis for commutative cone $\Tilde{\Gamma}$ and let $G = \sum c_k A_k$ be an arbitrary linear form in  $\Tilde{\Gamma}$.   Any $G\in \Tilde{\Gamma}$ decomposes as 
\begin{equation}
    \langle \{ \ket{a^k_p} \mid G \ket{a^k_p} = 0 \} \rangle \bigoplus \left(\ker G\right)^\perp.  
\end{equation} 
Then 
\begin{equation}\label{eq:kern}
    \langle \{ \ket{a^k_p} \mid G \ket{a^k_p} = 0 \} \rangle = \langle \left\{ \sum_{p,k} a_p^k \ket{a^k_p} \mid \sum_{p,k} a_p^k \ket{a^k_p} = 0 \right\} \rangle 
\end{equation}
and we say that the non-trivial solutions to $\sum_{p,k} a_p^k \ket{a^k_p} = 0$ define the union of the kernels of the operators in the sum defining $G$.
\end{definition}

We will now consider a special subclass of commutative cones that are represented by diagonal matrices acting on $[\mathbb{C}_d]^{\otimes}$.  We make a distinction from some literature \cite{CHH+89} and treat these cones as a subclass---the commutative penalty functions used in quantum annealing which arise in statistical mechanics in terms of generalized Ising models are called diagonal cones here. 

For diagonal $\otimes$-cones linear independence of Equation \eqref{eq:kern} simplifies since the basis elements now all share the same basis and satisfy a delta function relation.   

\begin{definition}\label{def:kernelalgebra1}
    Let $\Gamma_l\in \Tilde{\Gamma}$, for $\Tilde{\Gamma}$ a diagonal $\otimes$-cone, then the union of kernels is given as:  
\begin{equation}
    \ker\{\Gamma_l\} \cup \ker\{\Gamma_k\} = \Span_\mathbb{C} \{ \ket{{\bf{b}}} |~ {\bf{b}} \in \{1:s\}^{\times d},~ \Gamma_l  \ket{{\bf{b}}} =0 \vee \Gamma_k  \ket{{\bf{b}}} =0  \}, 
\end{equation}
and the intersection of kernels as,
\begin{equation}
    \ker\{\Gamma_l\} \cap \ker\{\Gamma_k\} = \Span_\mathbb{C} \{ \ket{{\bf{b}}} |~ {\bf{b}} \in \{1:s\}^{\times d},~ \Gamma_l  \ket{{\bf{b}}} =0 \wedge \Gamma_k  \ket{{\bf{b}}} =0  \}. 
\end{equation}
\end{definition}

\begin{definition}[Kernel algebra]\label{def:kernelalgebra2}
Let $\Tilde{\Gamma}$ be an diagonal $\otimes$-cone.  For any $\Gamma_1, \Gamma_2\in \Tilde{\Gamma}$ we have the following. 
\begin{enumerate}
    \item Sum to intersection rule: 
    \begin{equation}
    \ker\{\Gamma_1 + \Gamma_2 \} = \ker\{\Gamma_1\} \cap \ker\{\Gamma_2\}. 
    \end{equation} 
    \item Product to union rule: 
    \begin{equation}
    \ker\{\Gamma_1 \cdot \Gamma_2 \} = \ker\{\Gamma_1\} \cup \ker\{\Gamma_2\}. 
    \end{equation} 
\end{enumerate}
\end{definition}

These rules from Definition \ref{def:kernelalgebra2} allow one to form composition laws when engineering penalty functions of diagonal $\otimes$-comes.  We consider this in some detail in \S~\ref{sec:composition}.  

% \begin{example}
% Example of an commutative tensor cone,
% \begin{equation}
% \Tilde{B} = \Span_\mathbb{R}\{ M\in \diag\Mat_\mathbb{C}(s^d) |~ M \geq 0\}. 
% \end{equation}
% \end{example}

% Now we look at kernels of elements of commutative tensor cones.

% We rewrite the definition of the kernels of elements of commutative $\otimes$-cone as,

% \begin{equation}
% \ker\{ \Gamma_l \}= \Span_\mathbb{C}\{ \ket{\bf{x}} |~ \bf{x} \in  \{ 1:s \}^{\times d},~ \Gamma_l \ket{\bf{x}} = 0\}. 

% \begin{definition}[commutative matrix monoid]
% An commutative monoid is a set of matrices $G$ such that, $\forall l,~q_l \in G$, $[q_l, q_k]=0$,

% \begin{equation}
% \Gamma_l= \Span_{\mathbb{R}^+}\{ q_l \in G |~ q_l\geq 0 \}
% \end{equation}
% where $G$ is an commutative matrix monoid over $\mathcal{L}([\mathbb{C}_s]^{\otimes d})$.

% \begin{example}
% Let $S$ and $P$ be two sets. Using set builder notation define $S\cup P$ and $S\cap P$ on their elements. 
% \begin{equation}
%     S\cup P = \{ q|~ q\in S \vee q\in P \}
% \end{equation}
% \begin{equation}
%     S\cap P = \{ q|~ q\in S \cone q\in P \}
% \end{equation}
% \end{example}

% \end{definition}
%  Now we use this to define the union and intersection of kernels of commutative $\otimes$-cone.

% \begin{remark}\label{rem: Kerofsum-cone}
% \begin{equation}
%     \ker\{\Gamma_l + \Gamma_k \} = \ker\{\Gamma_l\} \cap \ker\{\Gamma_k\}
% \end{equation}
% \begin{equation}
%     \ker\{\Gamma_l \cdot \Gamma_k \} = \ker\{\Gamma_l\} \cup \ker\{\Gamma_k\}
% \end{equation}
% \end{remark}

\subsection{Pseudo Boolean algebras and the generalized Ising model}\label{sec:pseduo}

The generalized quadratic Ising model on $n$-qubits for $h_l, J_{lk} \in \mathbb{R}$ is given as: 
\begin{equation}\label{eq:ising}
    H = \sum_l h_l Z_l + \sum_{l\neq k}J_{lk} Z_l Z_k.  
\end{equation}
Equation \eqref{eq:ising} is written in the $\{\pm 1\}$ eigenbasis of the individual Pauli $Z$ operators.  The relationship between variables $z\in \{\pm 1\}$ and $x\in \{0, 1\}$ is: 
\begin{equation}\label{eq:z2x}
    z = 1-2x = (-1)^x. 
\end{equation}
We use lowercase $z$ to be a variable and $Z$ to be its matrix embedding.  Then Boolean values map to vectors $\ket{0}$ and $\ket{1}$ and $z$ maps to an eigenvalue as: 
\begin{equation}
    Z\ket{x} = (-1)^x \ket{x} = z\ket{x}, 
\end{equation}
where the last step assumes Equation \eqref{eq:z2x}. Returning to Equation \eqref{eq:ising} and applying Equation \eqref{eq:z2x} is given by an affine transformation on terms linear in $Z$: 
\begin{equation}
    \sum_l h_l Z_l \mapsto \sum_l h_l - 2 \sum_l h_l x_l.  
\end{equation}
However, when considering quardratic terms the mapping becomes: 
\begin{equation}
    \sum_l J_{lm} Z_l Z_m \mapsto \sum_{lm}J_{lm} - 2 \sum_{lm}(x_l + x_m) + 4 \sum_{lm}J_{lm}x_l x_m. 
\end{equation} 
Returning now to the vector space embedding.  We define the basis dependent map 
\begin{equation}
    f({\bf x}) \mapsto \sum_{{\bf x}\in \mathbb{B}^n} f({\bf x})\ket{{\bf x}}\bra{{\bf x}} = H_f.   
\end{equation}
Alternatively we could have sent $f({\bf x})$ to $\sum_{{\bf x}\in \mathbb{B}^n} f({\bf x})\ket{{\bf x}} = \ket{\psi_f}$.  These maps are related as: 
\begin{equation}
    H_f \ket{+^n} 2^{n/2} = \ket{\psi_f}. 
\end{equation} 

% \begin{remark}[Zero functions]
%     Note that the zero function is typically not considered.  
% \end{remark}

From these properties above, we arrive at the following commutative diagram: 
\begin{center}
    \begin{tikzcd}[column sep=large, row sep=huge]
        f({\bf x}) \arrow[rd] \arrow[r] & H_f \in \text{herm}_\mathbb{C}(2^n),\bra{{\bf x}} H_f \ket{{\bf x}} = f({\bf x}) \arrow[d] \\
        & \ket{\psi_f} \in [\mathbb{C}_2]^{\otimes n}, \braket{{\bf x}}{\psi_f}=f({\bf x}) 
    \end{tikzcd} 
\end{center}
Given $H_f$ or $\ket{\psi_f}$, we could recover a function $f'$ such that $\forall {\bf y} \in \mathbb{B}^n$, $f'({\bf y})=f({\bf y})$.   Then the original form of $f$ would typically be lost. 

\begin{remark}[Notation]
    We sometimes use falling powers notation.  Let $\bf x$ and $\bf \sigma$ be bit string variables.  Then, 
    \begin{equation}
        {\bf x}^{\bf \sigma} = x_1^{\sigma_1} \cdot x_2^{\sigma_2}\cdots x_n^{\sigma_n} 
    \end{equation}
    and the product of ${\bf x}^{\bf \tau}$ and ${\bf x}^{\bf \sigma}$ vanish iff ${\bf \sigma}\neq {\bf \tau}$ as 
    \begin{equation}
        {\bf x}^{\bf \sigma}{\bf x}^{\bf \tau} = \delta_{{\bf \sigma}, {\bf \tau}}. 
    \end{equation}
\end{remark}

We now recall the two canonical expansions. The  pseudo-Boolean functions can be expressed as a sum of disjoint variable products:
\begin{equation}\label{eqn:canonical}
    f({\bf x}) = \sum_{{\bf \sigma}\in \mathbb{B}} a_{\bf \sigma}{\bf x}^{\bf \sigma} = a_0 \overline{x}_1 \dots \overline{x}_n + a_1 x_1 \overline{x}_2\dots \overline{x}_n+\dots + a_N x_1\dots x_n.
\end{equation}
We use ${\bf x}$ to represent the variable vector $(x_1, x_2, \dots, x_n)$, $\forall l, a_l\in \mathbb{R}$, concatenated variables such as $x_k x_m$ are multiplied with the multiplication symbol omitted and finally, $\overline{x}$ is the logical negation of Boolean variable $x$.  We can replace all $\overline{x}\mapsto 1-x$ in Equation \eqref{eqn:canonical}.  After some calculation, one arrives at the canonical form, 
\begin{equation*}\label{eqn:canonical2}
\begin{split}
        f({\bf x}) = \sum_{S\subseteq [n]}C_S \prod_{k\in S} x_k =      
        c_0 + c_1 x_1 + c_2x_2 +\dots +c_n x_n +   \\
         c_{1,2}x_1x_2+\dots +c_{n-1, n}x_{n-1}x_n + \\
         \vdots \\ 
          +c_{1, 2,\dots n}x_1x_2\dots x_n. 
\end{split}
\end{equation*}
These forms, Equation \eqref{eqn:canonical}, Equation \eqref{eqn:canonical2} each uniquely define any $n$-variable pseudo-Boolean function in terms of $N=2^n$ real numbers $\bf a$ and $\bf c$.   If for any input ${\bf y}\in \{0,1\}$, $f(\bf y)\geq 0$ then $f(\bf x)$ is called non-negative.

We will define the kernel of a pseudo Boolean function to match the kernel of the corresponding operator through linear extension/flattening:

\begin{definition}[Kernel of a pseudo Boolean function]
    Let \(f({\bf x})\) be a pseudo Boolean function. The kernel of \(f\) is the set
    \begin{equation}
        \ker f({\bf x}) = \{{\bf x} \in \mathbb{B}^n \,|\, f({\bf x}) = 0\}.
    \end{equation}
    
\end{definition}

We note that the kernels of the pseudo Boolean function and its qubit embedding are related through linear extension (flattening) and 
    \begin{equation}
        \langle \ker f({\bf x}) \rangle = \ker H_f. 
    \end{equation}

\subsection{Parent Hamiltonian classification}\label{sec:classification}

We begin by considering a general 1-body penalty Hamiltonian:  
\begin{equation}\label{eq:G}
    g = c_0 + \sum_{k=1}^n c_k x_k^{\tau (k)}. 
\end{equation}
Here $\forall k, c_k\in \mathbb{R}_+$ and $\tau (k)$ is a binary variable that selects if $x_k$ appears in complimented ($x_k^0\mapsto 1-x_k$) or uncomplimented ($x_k^1\mapsto x_k$) form. 
The possible assignments of $n$ numbers in $\mathbb{R}_+$ and $n$ Boolean variables defining $\mathbf{\tau}$ in Equation \eqref{eq:G} parameterizes the space of 1-body non-negative penalty functions.  We are concerned with the minimisation of $g$ and arrive at 
\begin{equation}
    \arg\min g(x) = \arg \min \sum_{k=1}^n c_k x_k^{\tau (k)}
\end{equation}
and so the minimum assignment does not depend on the constant (energy) shift $a_0$.  Likewise, each term $c_k$ either adds or removes energy and depends solely on $x_k$.  The class of possible ground states remains unchanged through the restriction $c_k\in \mathbb{R}_+$.  

For the purpose of analyzing the possible ground states we will relabel our bits with respect to the element wise negation of $\mathbf{\tau}$.  In the gauge defined by $\overline{\mathbf{\tau}}$, we arrive at the simpler expression: 
\begin{equation}
    g_{\overline{\mathbf{\tau}}} ({\mathbf{x}}) = \sum_{k=1}^n c_k x_k^{\tau (k)\oplus \overline{\mathbf{\tau}}(k)} = 
    \sum_k c_k x_k
\end{equation}
which always has $0^n$ in its kernel instead of $\overline{\mathbf{\tau}}$ and when all $c_k$ are non-zero, we recover a non-degenerate ground state.  If we let some $p\leq n$ of the $c_k's$ vanish then we end up with a kernel of the form 
\begin{equation}
  \ker g_{\overline{\mathbf{\tau}}} ({\mathbf{x}}) \sim   \{\mathbb{B}^{p}, 0^{n-p}\}
\end{equation}
where we send elements raised to the power of zero to $\emptyset$ and $\mathbb{B}^{p}$ denotes $p$ free Boolean variables. We can use this to engineer ground states.  For example, suppose $\ker\{ f \}= \{ 0^k, 1^k \}$, then $f$ takes the form,
\begin{equation}\label{eq:ghzisingparent}
    f({\bf{x}})=\left[ \sum_k c_k x_k \right] \left[ \sum_k a_k -\sum_k a_k x_k \right]
\end{equation}
where $c_l, a_l \geq 0$ for all $l$.  Lifting the operator by linear extention, we arrive at a parent Hamiltonian for the $n$-qubit GHZ state: $\ket{0^n} + \ket{1^n} \in \langle \{\ket{0^n}, \ket{1^n}\} \rangle$.  Similar to the Sherrington-Kirkpatrick model \cite{Sherrington1975}, this construction typically necessitates all-to-all interactions.

From this construction we can derive an elementary toy model that has nice calculational properties and cryptographically protects the internal product structure through the difficulty of factoring.  Consider: 
\begin{equation}
    f({\bf x}) = \sum_k c_k x_k^{\pi(k)}\geq 0 
\end{equation}
and 
\begin{equation}
    g({\bf x}) = \sum_p c_p x_p^{\tau(k)}\geq 0. 
\end{equation}
For the case that the $c's$ are all equal to unity, we have that $\spec f({\bf x}) = [0:n]$.  We want to consider the product: 
\begin{equation}
    f({\bf x})\cdot g({\bf x}) = \sum_{k, p}c_k c_p x_k^{\pi(k)}x_p^{\tau(p)}.
\end{equation}
It follows that 
\begin{equation}
    \{\overline{\bf \pi}, \overline{\bf \tau}\} \in \ker f({\bf x})\cdot g({\bf x}) = \ker f({\bf x}) \cup \ker g({\bf x}).  
\end{equation}
To map this to a physical model with a 1D kernel we consider 
\begin{equation}
    \ker f^2 ({\bf x}) = \{ \overline{\bf \pi}\} 
\end{equation}
and 
\begin{equation}
    \spec f^2 ({\bf x}) = \{\lambda_{\bf k}^2 |  \lambda_{\bf k} = f({\bf k})\}.  
\end{equation}
We note that 
\begin{equation}
    x_k^{\sigma(k)}\mapsto \frac{1}{2}\left( \eye +(-1)^{\sigma(k)}Z_k\right). 
\end{equation}
Unlike the Sherrington-Kirkpatrick model, this model $f^2 ({\bf x})$ has 1-body terms.  A common choice for the Sherrington-Kirkpatrick model is to use a Gaussian distribution to determine the coupling coefficient, but choosing couplings from a simpler distribution, such as $\pm 1$, is also a valid approach. This binary distribution simplifies experimental implementation of the model while still capturing the essential features of disorder and frustration that are characteristic of spin glasses.  These properties are also present in $f^2 ({\bf x})$.  Note that the origonal proof of the NP-hardness of the Ising model considering couplings $\{0, \pm 1\}$ on a square lattice \cite{Barahona1982}.  

% We note that
% \begin{equation}
%     \overline{x}_k\mapsto \frac{1}{2}\left( \eye -Z_k\right) 
% \end{equation}
% \begin{equation}
%     x_k\mapsto \frac{1}{2}\left( \eye +Z_k\right) 
% \end{equation}
% and we arrive at local terms and a quardratic pairing that distributes the binary variable $\pm 1$ as 
% \begin{equation}\label{eqn:skrestricted}
%    \propto \sum_{k, p} (-1)^{\sigma(k)+\sigma(p)} Z_kZ_p.  
% \end{equation}
% Unlike the Sherrington-Kirkpatrick model, this model has 1-body terms.  However, a common choice for the Sherrington-Kirkpatrick model is to use a Gaussian distribution to determine the coupling coefficient, but choosing couplings from a simpler distribution, such as $\pm 1$ in Equation \eqref{eqn:skrestricted}, is also a valid approach. This binary distribution simplifies experimental implementation of the model while still capturing the essential features of disorder and frustration that are characteristic of spin glasses.  

\subsection{Root systems and factorisations}
\label{sec:roots}

We will now consider the case of symmetric penalty functions, which can be factored over $\mathbb C$.  By $\omega{\bf (x)}:\mathbb{B}^n \to \mathbb{B}^n$  we denote a permutation of the elements of the tuple ${\bf x} \in \mathbb{B}^n.$

\begin{definition}\label{def:sympbf}
A pseudo-Boolean function $f:\{0,1\}^n\to \mathbb{R}$ is called symmetric if for all ${\bf x} \in \mathbb{B}^n \ f({\bf x})=f(\omega{\bf (x)})$ for all $\omega{\bf (x)}.$ 
\end{definition}

\begin{remark}
 The authors \cite{ABCG16} state equivalently that, a symmetric pseudo-Boolean function $f:\{0,1\}^n\mapsto \mathbb{R}$ depends only on the Hamming weight of the input.  The Hamming weight of a bit vector ${\bf x}$ is given by the 1-norm
    \begin{equation}
        \|{\bf x}\|_1 = \sum_{l=1}^n x_l
    \end{equation}
and hence, a pseudo-Boolean function is symmetric whenever there exists a discrete function $k:\{0, 1, \dots, n\}\mapsto \mathbb{R}$ such that $f({\bf x}) = k(\|{\bf x}\|_1)$. 
\end{remark}

\begin{remark}
From Equation \eqref{eqn:canonical}, it follows that the canonical form of a symmetric pseudo-Boolean function is  
\begin{align}\label{eq:canonspb}
    f({\boldsymbol {x}}) = & a_0 + a_1\sum _{k_1}x_{k_1} + a_{2}\sum_{k_1<k_2}x_{k_1}x_{k_2} + a_{3}\sum _{k_1<k_2<k_3}x_{k_1}x_{k_2}x_{k_3} + \ldots \nonumber \\
    & + a_{n}\sum _{k_1<k_2<k_3< \ldots <k_n}x_{k_1}x_{k_2}x_{k_3}\ldots x_{k_n}. 
\end{align}

\end{remark}

Beginning with a series expansion, we now derive some results specific for symmetric pseudo-Boolean functions.  

%By ${j \brace i}$ 
\begin{remark}[Notation]
By $\{\substack{j \\ i}\}$ we denote the Stirling number of the second kind i.e.~the number of ways to partition a set of $j$ elements into $i$ nonempty subsets.  
\end{remark}

\begin{theorem}[Series expansion---Biamonte and Sengupta \cite{SB23}]\label{thm:series}
Let  $f:\mathbb{B}^n\to \mathbb{R}$ be a pseudo-Boolean function in canonical form Equation \eqref{eq:canonspb}, ${\bf a}=(a_1,a_2,\ldots,a_n)$  and $c_0=a_0$ then there exists a unique ${\bf c}=(c_1,c_2,\ldots,c_n)\in \mathbb{R}^n$ and such that 
\begin{equation}\label{summ}
f({\bf x})=\sum_{l=0}^nc_l\left(\sum_{k=1}^{n}x_k\right)^l,
\end{equation}
 \begin{equation}
   {\bf a}= B{\bf c},
 \end{equation}
 where 
 \begin{equation}
     B =\displaystyle \begin{pmatrix}
1! \{\substack{1 \\ 1}\} & 1! \{\substack{2 \\ 1}\} & 1! \{\substack{3 \\ 1}\} & \dotsc  & 1! \{\substack{n \\ 1}\}\\\\
0 & 2! \{\substack{2 \\ 2}\} & 2! \{\substack{3 \\ 2}\} & \dotsc  & 2! \{\substack{n \\ 2}\}\\\\
0 & 0 & 3! \{\substack{3 \\ 3}\} & \dotsc  & 3! \{\substack{n \\ 3}\}\\
\vdots  & \ddots  & \ddots  & \ddots  & \vdots\\
0 & 0 & \ldots & 0 & n! \{\substack{n \\ n}\}
\end{pmatrix}.
\end{equation}
\end{theorem}

\begin{corollary}[Product factorisation]
 A symmetric pseudo-Boolean function $f:\mathbb{B}^n\to \mathbb{R}$ of degree $n$   can be expressed as
 \begin{equation}\label{fact}
      f({\bf x})=K\prod_{l=1}^n\left(\lambda_l-\sum_{k=1}^{n}x_k\right), ~~\lambda_l, K \in \mathbb{C}.
\end{equation}
\begin{proof}
 Substituting $\sum_{k=1}^{n}x_k=X$ into (\ref{summ}) we obtain a polynomial $Q(X)$ of degree $n$ in one variable. Since, $\mathbb{C}$ is algebraically closed, this polynomial will have $n$ roots $\lambda_l \in \mathbb{C}.$ 
\end{proof}
\end{corollary}

In equation (\ref{fact}) even though $\lambda_l \in \mathbb{C},$ only real values are obtained when $x_i\in \mathbb{B}$  which matches with the corresponding values of the unfactored symmetric pseudo-Boolean function. 
We note that $f({\bf x})$ vanishes identically for inputs constrained under Equation \eqref{fact} as 
\begin{equation}
   \sum_{k=1}^{n}x_k =  \lambda, 
\end{equation}
defining the kernel of $f({\bf x})$. The matrix embedding evaluates the function only at Boolean entries.  In terms of optimisation, this looses nothing as the following theorem by Boros and Prékopa \cite{BP89}, Rosenberg \cite{Ros72} and reviewed by Boros and Hammer \cite{BH02} shows: 
\begin{theorem}[Extrema at Boolean inputs~\cite{BP89, Ros72}]\label{thm:extrema}
Consider a pseudo-Boolean function $f(x_1,x_2,\ldots,x_n)$ and let ${\bf r} \in [0,1]^{n}$. Then there exist binary vectors
$ {\bf x}; {\bf y} \in \{0,1\}^n$ for which
$f({\bf x}) \leq f({\bf r}) \leq f({\bf y})$. 
\end{theorem}

\subsection{Examples}
\label{sec:examples} 
The following examples realize specific functions in terms of polynomials.  Ground state embedding will be considered in \S~\ref{sec:composition}.  Here we will expand $\delta_{l m n}$ as a pseudo-Boolean function in the evident way, as the sum of two disjoint terms: 
\begin{equation}\label{delt}
    \delta_{k l m} = k l m  + (1-l)(1-m)(1-k) = 1 - l - m - k + m k + kl + lm. 
\end{equation}
From Theorem \ref{thm:series} this can be equivalently expressed as 
\begin{equation}
\begin{split}\label{delt2}
        \delta_{l m k} = \frac{1}{2} (l + m + k)^2 - \frac{3}{2}(l + m + k) + 1   \\
        =\frac{1}{2} (l + m + k -2) (l + m + k -1).
\end{split}
\end{equation}
We will see that the root system $\{1, 2\}$ provides a classification of (\ref{delt2}) where the function vanishes identically whenever 
\begin{equation}
    \begin{split}\label{eqn:delker} 
        l + m + k = 2, \\
        l + m + k = 1. 
    \end{split}
\end{equation}
Together the Equations \eqref{eqn:delker} have six solutions over $\{0,1\}^3$ whereas the entire solution space corresponds to the two hyperplanes  $l+m+k=1,2$.  

Likewise,  one would express the $\sf XOR$ function as 
\begin{equation} 
\begin{aligned}
  \text{XOR}(l, m, k) &= l\oplus m\oplus k = \frac{2}{3}(l + m + k)^3-3(l + m + k)^2 +\frac{10}{3}(l + m + k) \\
  &= \frac{2}{3}(l + m + k - 2)(l + m + k )(l + m + k - \frac{5}{2}).
  \end{aligned}
\end{equation} 
This example has fractional roots. These occur whenever a root (such as $5/2$) appears outside of the Boolean input domain. In this case, the kernel over the Boolean domain $\{0,1\}^3$ corresponds to only four elements, whereas all solutions correspond to the three hyperplanes $l+m+k=0, 2, 5/2$. 

We can write the delta function on $k$ binary variables (denoted $\bf m$) in its product form viz.,  
\begin{equation}
    \delta_{\bf m} = \frac{(-1)^{k
    -1}}{ (k-1)!} \prod_{l=1}^{k-1} \left[\sum_{l=1}^k x_l - l\right]. 
\end{equation}

Here, $\delta_{\bf m}$ has $k-1$ roots $\{1,2,3,\ldots,k-1\}$ as the symmetric form is of degree $k-1$ and with the kernel corresponding to $k$ long bit vectors  from $\{0,1\}^n\setminus \{0^{\times k},1^{\times k}\}.$

It can be obtained that the symmetric form of the XOR function  of $n$-variables has the form 
\begin{equation}\label{XOR symm}\small
\begin{split}
    f({\boldsymbol {x}})=\sum _{i_1}x_{i_1} -2\sum_{i_1<i_2}x_{i_1}x_{i_2}+4\sum _{i_1<i_2<i_3}x_{i_1}x_{i_2}x_{i_3}+\ldots\\
     + (-2)^n\sum _{i_1<i_2<i_3< \ldots <i_n}x_{i_1}x_{i_2}x_{i_3}\ldots x_{i_n}. 
\end{split}
\end{equation}
One might factor the symmetric form using  Theorem \ref{thm:series} but the roots obtained are not always aesthetically pleasing.

Symmetric pseudo-Boolean functions also arise when considering the symmetric Ising model from statistical and quantum mechanics: 
\begin{equation}\label{eqn:ham2}
    H({\bf x}) = \frac{J}{2} \sum_{l\neq m} x_lx_m  + h \sum_l x_l.  
\end{equation}
The quantity $\sum_{l=1}^n x_l$ corresponds to the total spin along the Z-axis for any state of the system, the non-zero coupling strength $J$ sets the interaction strength and $h$ is called the local bias. The symmetric Ising model is typically considered with variables $z = \pm1$ and with the affine transformation ($z=1-2x$) can be written up to a constant as follows
\begin{equation}\label{eqn:ham}
    H({\bf x}) = \frac{J}{2} \sum_{l\neq m} x_lx_m  + h \sum_l x_l. 
\end{equation}
In (\ref{eqn:ham}) the coupling strength $J$ is non-zero (otherwise Equation \eqref{eqn:ham} is trivial) and the bias $h$ takes only real values. Using our methods, Equation \eqref{eqn:ham} is equivalently expressed as 
\begin{equation}
    H({\bf x}) = \frac{J}{4}\left(\sum_{l=1}^n x_l + \frac{4h}{J} - 1\right) \sum_{l=1}^n x_l.   
\end{equation}
The quantity $\sum_{l=1}^n x_l$ corresponds to the total $Z$-spin for any state of the system and its expected value vanishes whenever,
\begin{equation}
    \begin{split}
        \sum_{l=1}^n x_l = 0,~~ \sum_{l=1}^n x_l = 1-\frac{4h}{J}, \\
            \end{split}
\end{equation}
and hence,  states restricted to either of these two hyperplanes correspond to the kernel of $H({\bf x})$.

% In general, given $\bf{v}$ the penalty function,

% \begin{equation}
% f_{\bf{v}}(\bf{x}) = \left[\sum_{k=1}^n c_k x_k^{\overline{v}_k} \right]^m
% \end{equation}
% has $\ker\{ f(\bf{v}) \} = \Span_\mathbb{C}\{ \ket{\bf{v}} \}$ for all $k$, $c_k\geq 0$. Where $1\leq m\leq n$, and $n$ is the number of variables.

% In general the operator,

% \begin{equation}
% \Tilde{Q} = \prod_{l=1,~ v_1\dots v_p}^p \left[\sum_{k=1}^n c_k x_k^{\overline{v}_{l(k)}} \right]^{m_l}
% \end{equation}
% has $\ker\{ \Tilde{Q} \}= \Span_\mathbb{C}\{ \bf{v}_1, \cdots, \bf{v}_p \}$, for all $l$, $m_l \in \mathbb{N}$.

\section{Constructing parent Hamiltonian's by function composition}\label{sec:composition}

\subsection{Function composition}

We draw composition of functions by connecting output wires to inputs,  depicted graphically as follows:
\begin{equation}\label{eq:f-comp}
    \includegraphics[]{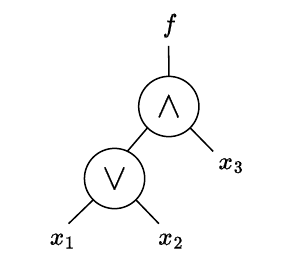}
\end{equation}
We write Equation \eqref{eq:f-comp} as,
\begin{equation}\label{eq: fun-a-o}
f(x_1, x_2, x_3) = \wedge (\vee(x_1, x_2), x_3).  
\end{equation}
We can express Equation \eqref{eq:f-comp} as a polynomial: $\wedge (\vee(x_1, x_2), x_3) = (x_1+x_2-x_1x_2)x_3$.  We are instead interested in embedding $f$ into an operator kernel.  Let us consider the following diagram: 
\begin{equation}
    \includegraphics[]{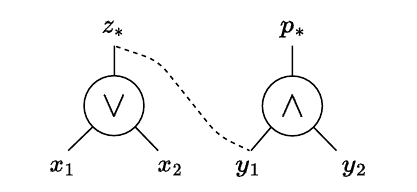}
\end{equation}
where the dashed line is contracting  the upper leg of the OR with the left leg of the AND, which we write as,
\begin{equation}
f_\vee(x_1, x_2, z_*) + f_\wedge(z_*, y_2, p_*). 
\end{equation}
We will write the kernel of the $\vee$-gate as 
\begin{equation}
    \ker\{ f_\vee \} = \{ (x_1, x_2, z_\star)|~ z_\star= x_1\vee x_2 \}
\end{equation}
and the $\wedge$-gate as 
\begin{equation}
    \ker\{ f_\wedge \} = \{ (y_1, y_2, p_\star)|~ p_\star= y_1\wedge y_2 \}. 
\end{equation}
From the kernel algebra we have that 
\begin{equation}
    \ker\{ f_\vee + f_\wedge \} = \{ (x_1, x_2, y_2, p_\star) |~ p_\star= y_1\wedge z_\star, z_\star= x_1\vee x_2 \}
\end{equation}
where we have omitted the intermediate variables $z_\star$ and $y_1$ in the input set $(x_1, x_2, y_1, p_\star)$.  In practice the penalty function must be minimized.   Composition of kernel embedding is done by equating variables as such it preserve locality at the cost of adding slack bits.  
In terms of applications in practice, any Boolean formula can be embedded in this fashion.  We then want to add a penalty to the output forcing the output bit $p_\star$. 
Consider
\begin{equation}\label{eq:andor}
\mathop{\arg\min}_{x_1, x_2, y_2} f_\vee (x_1, x_2, z_*) + f_\wedge(z_*, y_2, 1)= \{(x, y, z) |~ \wedge (\vee(x, y), z)= 1\}. 
\end{equation}
Minimization of Equation \eqref{eq:andor} determines an input to cause the function composition representing $\wedge (\vee(x_1, x_2), x_3)$ to output 1. 
Arising in adiabatic quantum computing \cite{Biamonte2008} related situation arises in tensor networks \cite{Biamonte2015}.  Consider the following Boolean tensor network embedding $f: \mathbb{B}^n\rightarrow \mathbb{B}$,
\begin{equation}
    \includegraphics[]{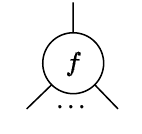}
\end{equation}
then by bending the top wire down we arrive at the following state,
\begin{equation}\label{eq: f-to-state}
    \includegraphics[]{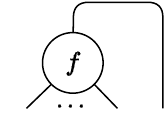}
\end{equation}
which is written as,
\begin{equation}
    \ket{\psi} = \sum_{\bf{x}} \ket{\bf{x}, f(\bf{x})}
\end{equation}
We write the following diagram as state, 
\begin{equation}\label{eq: ex-state}
    \includegraphics[]{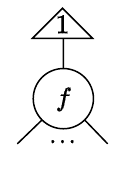}
\end{equation}
and arrive at,
\begin{equation}
    \begin{aligned}
        \sum_{\bf{x}} \eye \otimes \bra{1} \circ \ket{\bf{x}} \ket{f(\bf{x})} &=  \sum_{\bf{x}} \eye \ket{\bf{x}} \otimes \braket{1}{f(\bf{x})} \\
        &= \sum_{\bf{x}} f(\bf{x}) \ket{\bf{x}}
    \end{aligned}
\end{equation}
and so any of these bit strings satisfy $f$ provided there is at least one bit string that satisfies $f$, otherwise  $\sum_{\bf{x}} f({\bf{x}}) \ket{{\bf{x}}}=0$.

The kernel algebra of a diagonal $\otimes$-cone provides a mechanism to engineer ground states.  The primary question is to determine an Ising model that is minimized on a given set of bit strings, formally: 

\begin{definition}[Ising Kernel Problem]
Given $S\subseteq \{0,1\}^{\times n}$ where $1\leq |S|\leq 2^n$. Does there exist a non-negative diagonal $\otimes$-cone such that 
\begin{equation}
    \ker \{h \} = \Span \{ \ket{s}| s \in S\}?
\end{equation}
\end{definition} 

There are several scenarios where this issue will arise. The inverse Ising problem aims to approximate solutions to this question on large-scale lattices \cite{NZB17}. Additionally, one might consider a quantum state on qubits and develop a parent Hamiltonian construction based on it. 

\begin{definition}[Support of $\ket{\psi}$]
    Let $\ket{\psi} = \sum_{x\in \mathbb{B}^n} c_x \ket{x}$ then 
    \begin{equation}
        \lfloor \ket{\psi} \rfloor \mapsto \{ x|\braket{x}{\psi}\neq 0\}_x 
    \end{equation}
    is called the support of $\ket{\psi}$ and $\lfloor \ket{\psi} \rfloor$ is called $\ket{\psi}$'s flattening.  
\end{definition}

From the support map we arrive directly at a parent Hamiltonian for $\ket{\psi}$.  The construction is degenerate though still has some theoretical interest.  We will now consider a construction to introduce slack bits to keep two-body locality.  

\begin{definition}[Kernel embedding]
A kernel embedding of a function $f(\bf{x}):\mathbb{B}^n \rightarrow \mathbb{B}$ is given by a function $g(\bf{x}, \bf{y}):\mathbb{B}^{n+1}\rightarrow \mathbb{R}_+$ such that
\begin{equation}
 \ker\{g(\bf{x}, \bf{y})\}= \{ (\bf{x}, f(\bf{x}))|~ \bf{x}\in \mathbb{B}^n \}   
\end{equation}
where $g(\bf{x}, \bf{y})$ embeds the truth table into the kernel. 
\end{definition}

Consider $\arg\min_{{\bf{y}}\in \mathbb{B}^n} g(\bf{y}, 1)$ which is equal to $\{ \bf{x}|~ f(\bf{x})=1 \}$.  Let $\Tilde{\Gamma}$ be a $3$-SAT instance. Let $\Tilde{P}$ embed $\Tilde{\Gamma}$ in its kernel. So $\Tilde{P}\geq 0$, and $\ker\{ \Tilde{P} \}= \{ \bf{x}|~ \Tilde{\Gamma}(\bf{x}) = 1 \}$. $\Tilde{\Gamma}$ is satisfiable if and only if $\ker\{\Tilde{P}\} \neq \emptyset$.

\section{Conclusion}

In this work, we considered engineering Hamiltonian ground states within a commutative cone, blending techniques from different disciplines. We employed several techniques that cut across traditional discipline boundaries and presented a unified framework which expresses aspects of the contemporary theory of commutative parent Hamiltonians.  Our constructions don't offer the ideal physical property of spatial locality: indeed, in the non-degenerate case even $n$-body terms arise.  For example, whereas our GHZ construction has a unique ground state the MPS construction is degenerate and yields a quadratic Ising operator \cite{Per+06} .  Our construction is non-local but also non-degenerate.  Our second construction, like non-invective MPS \cite{Per+06, Fern_ndez_Gonz_lez_2014} , results in a degenerate parent Hamiltonian for GHZ.  The construction and some of the results have an expository flavor and the framework and the results can serve the basis for a better understanding of the types of parent Hamiltonians that arise as penalty functions in quantum computing.

\section*{Acknowledgements}
{    
The author declares no competing interests. We thank Stuart Hadfield, Aaron Lott and Erik Gustafson for reading this paper and providing feedback.  Part of this work is funded by U.S.~Department of Energy, Office of Science, National Quantum Information Science Research Centers, Co-Design Center for Quantum Advantage under Contract No.~DE-SC0012704 through the NASA-DOE interagency agreement SAA2-403601.  JB was supported by NASA Academic Mission Services, Contract No.~NNA16BD14C.
}

\bibliographystyle{ieeetr}
\bibliography{bib}
\end{document}